\documentclass[aps,pra,twocolumn,floatfix,superscriptaddress,showpacs,psfig,reprint,longbilbiography]{revtex4-2}
\usepackage{amsmath}
\usepackage{amssymb}
\usepackage{amsthm}
\usepackage{amsfonts}
\usepackage[version=4]{mhchem}
\usepackage{listings}
\usepackage{enumerate}
\usepackage{latexsym}
\usepackage{psfrag}
\usepackage{bm}
\usepackage{ulem}
\usepackage[all]{xy}
\usepackage{graphicx}
\usepackage{subfigure}
\usepackage[pdftex,colorlinks]{hyperref}
\usepackage{color}
\usepackage{mathtools}
\usepackage{times}
\usepackage{array}
\usepackage{placeins}
\usepackage{comment}
\usepackage{lipsum}
\hypersetup{colorlinks=true,linkcolor=blue,filecolor=blue,urlcolor=blue,citecolor=blue}

\begin{document}
\title{Emergent supercounterfluid and quantum phase diagram of two-component interacting bosons in one-dimensional optical lattice}

\author{Saisai He}
\affiliation{Lanzhou Center for Theoretical Physics, Lanzhou University, Lanzhou 730000, China}
\affiliation{Key Laboratory of Quantum Theory and Applications of MoE, Lanzhou University, Lanzhou 730000, China}
\affiliation{Key Laboratory of Theoretical Physics of Gansu Province$\&$Gansu Provincial Research Center for Basic Disciplines of Quantum Physics, Lanzhou University, Lanzhou 730000, China}

\author{Yang Liu}
\affiliation{Lanzhou Center for Theoretical Physics, Lanzhou University, Lanzhou 730000, China}
\affiliation{Key Laboratory of Quantum Theory and Applications of MoE, Lanzhou University, Lanzhou 730000, China}
\affiliation{Key Laboratory of Theoretical Physics of Gansu Province$\&$Gansu Provincial Research Center for Basic Disciplines of Quantum Physics, Lanzhou University, Lanzhou 730000, China}

\author{Bin Xi}
\affiliation{College of Physics Science and Technology, Yangzhou University, Yangzhou 225002, China}

\author{Hong-Gang Luo}
\affiliation{Lanzhou Center for Theoretical Physics, Lanzhou University, Lanzhou 730000, China}
\affiliation{Key Laboratory of Quantum Theory and Applications of MoE, Lanzhou University, Lanzhou 730000, China}
\affiliation{Key Laboratory of Theoretical Physics of Gansu Province$\&$Gansu Provincial Research Center for Basic Disciplines of Quantum Physics, Lanzhou University, Lanzhou 730000, China}

\author{Qiang Luo}
\email[]{qiangluo@nuaa.edu.cn}
\affiliation{College of Physics, Nanjing University of Aeronautics and Astronautics, Nanjing 211106, China}

\author{Jize Zhao}
\email{zhaojz@lzu.edu.cn}
\affiliation{Lanzhou Center for Theoretical Physics, Lanzhou University, Lanzhou 730000, China}
\affiliation{Key Laboratory of Quantum Theory and Applications of MoE, Lanzhou University, Lanzhou 730000, China}
\affiliation{Key Laboratory of Theoretical Physics of Gansu Province$\&$Gansu Provincial Research Center for Basic Disciplines of Quantum Physics, Lanzhou University, Lanzhou 730000, China}

\date{\today}

\begin{abstract}
Motivated by a recent experiment that realizes nearest-neighbor dipolar couplings in an optical lattice [C. Lagoin, \textit{et al.}, \href{https://doi.org/10.1038/s41586-022-05123-z}{Nature \textbf{609}, 485 (2022)}], we study a one-dimensional version of the two-component extended Bose-Hubbard model via the density-matrix renormalization group method.
By using the nearest-neighbor and on-site interaction parameters from the experiment, we start by mapping the quantum phase diagram in the hopping parameters $t_{A}\mbox{-}t_{B}$ plane
with boson densities $\rho_{A}=\rho_{B}=1/2$.
In addition to the density wave phase reported in the experiment, we find several regimes of superfluidity when one or two hopping parameters are large enough, and interestingly there is a supercounterfluid phase at moderate and comparable hopping parameters.
The universality classes of these phase transitions are analyzed from the correlation functions, excitation gaps, and entanglement entropy. In particular, a Berezinskii-Kosterlitz-Thouless type is recognized for several gapped-to-gapless transitions.
In addition, we also study the quantum phase transitions when varying $\rho_{B}$ from 0 to 1 while keeping $\rho_A = 1/2$. 
We identify a supersolid phase in a wide range of $1/2<\rho_B<1$.
Our work paves the way for realizing exotic many-body phases in cold atom experiments upon proper tuning of experimental parameters.
\end{abstract}

\maketitle

\section{INTRODUCTION}
With the rapid development of atomic quantum simulators that operating at ultralow temperatures, ultracold atoms in optical lattices offer versatile platforms for understanding interacting quantum systems and explore alluring quantum phenomena~\cite{Lewenstein2007,Bloch2008}.
Particularly, they enable the simulation of a broad range of models exemplified by the Bose-Hubbard model~\cite{Fisher1989,Jaksch1998}.
In weakly interacting boson gases, s-wave scattering is capable of capturing the dominant low-energy physics~\cite{Bloch2008,Dutta2015}.
It is revealed that relevant systems can usually be well described by the Bose-Hubbard model, in which the superfluid~(SF) phase and Mott insulator~(MI) phase are reported \cite{Greiner2002}. Since dipole-dipole coupling in magnetic dipolar or Rydberg atoms can dominate contact interactions, long-range contributions generally cannot be neglected and may play a key role~\cite{Lahaye2009,Trefzger2011,Baranov2012}. Equipping the standard Bose-Hubbard model with further-range interactions naturally leads to the so-called extended Bose-Hubbard model~(EBHM). These long-range interactions, even with only the nearest-neighbor term, may modify the underlying phase diagram and induce novel phases.
Generally, the ground state varies from the SF phase for small on-site ($U$) and nearest-neighbor ($V$) interactions,
to insulating phases including the MI phase with large $U$ and the density-wave (DW) phase with large $V$ on the opposite side.
Upon tuning the interactions, the boson density, and the dimensionality, 
novel phases such as Haldane insulator~\cite{Dalla2006,Berg2008,Deng2011,Rossini2012,Xu2018}, supersolid~(SS)~\cite{Batrouni2006,Ejima2014,Mishra2009,Batrouni2014}, and phase-separated SF+SS~\cite{Kottmann2020,Kottmann2021} are reported.
Further, unconventional phase transitions such as the Berezinskii-Kosterlitz-Thouless (BKT) transition~\cite{Kuhner1998,Thamm2025} and deconfined quantum critical points \cite{Wang2025} are identified.

On the other hand, atoms trapped in optical lattices are not limited to single-component systems, but can also exist in multicomponent and spinor systems.
The use of elliptically polarized light has been proposed to tune interspecies interactions between the two hyperfine states of $\mathrm{^{87}Rb}$ atoms~\cite{Mandel2003}, for the realization of a two-component Bose-Hubbard model~\cite{Isacsson2005}. Alternatively, adjusting the layer spacing between two-component dipolar bosons in a bilayer optical lattice provides another possible route~\cite{Arguelles2007,Trefzger2009}. 
In addition to interactions within individual components, intercomponent on-site interaction is incorporated. Over the years, this model has been extensively examined by massive theoretical works~\cite{Kuklov2003,Kuklov2004a,Mathey2007,Mishra2007,Hu2009,Zhao2014}.
Competition between these interactions is essential
in stabilizing the paired superfluid~(PSF), supercounterfluid~(SCF), and phase separated MI phases~\cite{Kuklov2004a,Mishra2007,Hu2009}.
Quite recently, the SCF phase is realized in an ultracold atom experiment that mimics the two-component Bose-Hubbard model at unit fillings~\cite{Zheng2025}.
Additionally, theoretical work has suggested the possibility that equipping the two-component Bose-Hubbard model with long-range interactions in an array of optical cavities, such as nonlinear coupling terms between the components, can lead to much richer physical phenomena~\cite{Zhang2015}.

Recently, significant progress has been made experimentally in implementing long-range inter- and intracomponent interactions. By confining semiconductor dipolar excitons~(bosons)
in a two-dimensional optical lattice~\cite{Lagoin2022}, interacting bosons are localized within two Wannier states,
which can be effectively described by a two-component EBHM.
However, because of the inequivalence of two Wannier states, the model parameters for these two components are different. 
Although the two-component EBHM has been studied~\cite{Mishra2008,Zhang2022}, few works consider the inequivalent case~\cite{Mishra2010,Watanabe2024}.
Moreover, the mean-field theory based on the Gutzwiller ansatz~\cite{Ozaki2012,Machida2022} can provide some intuitive insights into the behavior of interacting bosons.
However, it may be inadequate in strongly interacting and critical regimes where quantum fluctuations are greatly enhanced although a modified version \cite{Caleffi2020,Victor2022} may partially compensate for these shortcomings.

To gain an intuitive understanding of the quantum phases and phase transitions which are born out of the competition between intra- and intercomponent interactions,
we employ the density-matrix renormalization group method~(DMRG)~\cite{White1992,Peschel1999,Schollwock2005} to examine the ground state of the two-component EBHM in one dimension. In the following, the two components are marked as $A$ and $B$, respectively. 
The density-density interactions reported in Ref.~\cite{Lagoin2022} are used in this work. 
In this work, we consider two cases. 
In one case, we fix the boson density of both components to half filling and adjust the hopping parameters to obtain a ground-state phase diagram.
In particular, we confirm the existence of DW phase reported in the experiment when the hopping parameters are small.
In the other case, starting from the DW phase at tiny hopping parameters and tuning the boson density of one component from half filling, we find that the system is then driven into a SS or SF state.

The remainder of this paper is organized as follows. The two-component EBHM and relevant observables to characterize the phase diagram are introduced in Sec.~\ref{sec:mod}, followed by a detailed numerical analysis in Sec.~\ref{sec:num}. 
Particularly, Sec.~\ref{SEC2:HalfFilling} presents the ground-state phase diagram of the two-component EBHM for both boson densities at half filling, with an emphasis on the SCF phase and its associated phase transitions.
By fixing hopping parameters $t_{A} = 2t_{B} = 0.04$ at which the ground state is the DW phase when $\rho_{A} = \rho_{B} = 1/2$, Sec.~\ref{sec:den} displays evolution of the quantum phases with respect to the boson density $\rho_B$ while keeping $\rho_A$ unchanged. Finally, Sec.~\ref{sec:con} is devoted to the conclusions.

\begin{figure*}[!t]
	\centering	
	\includegraphics[width=0.90\linewidth]{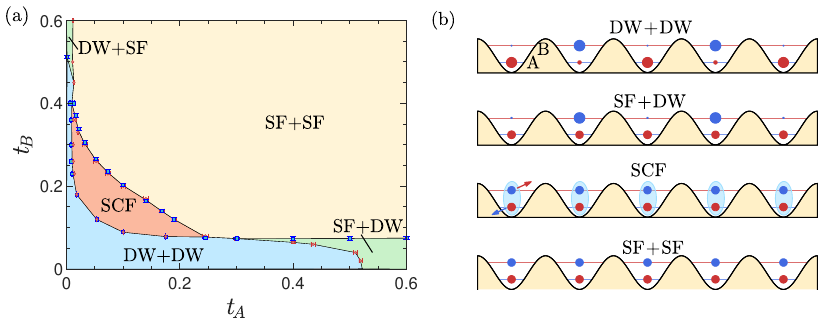}
	\caption{(a) Ground-state phase diagram of the two-component EBHM in the $t_{A}\mbox{-}t_{B}$ plane with the boson densities $\rho_{A}=\rho_{B}=1/2$. There are five different phases, marked as DW+DW, SF+DW, DW+SF, SCF and SF+SF. 
    The data points marked by red circles and blue squares establish the boundaries between these phases for component A and B bosons, respectively.
    (b) Sketch of the density distribution in the lattice for the phases shown in (a). The abbreviations DW, SF, and SCF indicate density wave, superfluid, and supercounterfluid, respectively.}
	\label{Fig:PhaseDiagram}
\end{figure*}

\section{MODEL and METHOD}\label{sec:mod}

In the following, we consider one-dimensional dipolar bosons confined in two Wannier states $A$ and $B$.
The low-energy behavior of such interacting bosons is described by the two-component EBHM whose Hamiltonian $\hat{H} = \hat{H}_{0} + \hat{H}_{C}$, in which
\begin{eqnarray}
  \hat{H}_{0} &\!=\!& \sum_{\langle ij\rangle\sigma}\big[-t_{\sigma}\big(\hat{b}^{\dagger}_{i\sigma}\notag
\hat{b}_{j\sigma}+{\text{h.c.}}\big)+V_{\sigma}\hat{n}_{i\sigma}\hat{n}_{j\sigma}\big]\notag\\
&&+\frac{U_{\sigma}}{2}\sum_{i\sigma}\hat{n}_{i\sigma}(\hat{n}_{i\sigma}-1), \\
\hat{H}_{C} &\!=\!& V_{C}\!\sum_{\langle ij\rangle}\big(\hat{n}_{iA}\hat{n}_{jB}+\hat{n}_{iB}
\hat{n}_{jA}\big) + U_{C}\!\sum_{i}\hat{n}_{iA}\hat{n}_{iB}.
\end{eqnarray}
Here, $\hat{b}^{\dagger}_{i\sigma}\,(\hat{b}_{i\sigma})$ is creation (annihilation) operator of the $\sigma(=A,B)$ component at the site $i$
and $\hat{n}_{i\sigma}\,$=$\,\hat{b}^{\dagger}_{i\sigma}\hat{b}_{i\sigma}$. $t_{\sigma}$ denotes the hopping parameter, and $U_{\sigma}$ and $U_{C}$ ($V_{\sigma}$ and $V_{C}$) represent intra- and intercomponent on-site (nearest-neighbor) interactions, respectively.
Following the experiment \cite{Lagoin2022}, values of these parameters are estimated as $t_A = 0.001$, $t_B = 0.007$, and 
\begin{subequations}\label{EQ:UabVab}
\begin{align}
&U_{A} = 1.0, U_{B} = 0.5, U_{C} = 0.2,  \\
&V_{A} = 0.035, V_{B} = 0.25, V_{C} = 0.04.
\end{align}
\end{subequations}
Here, the energy scale is set by choosing $U_A$ as the energy unit.
By virtue of the self-consistent mean field calculation, it is observed that the model exhibits two insulating phases, the DW phase featuring a checkerboard pattern and the MI phase, in a two-dimensional square lattice~\cite{Lagoin2022}.
These findings agree well with the experimental observations.
One may notice that the hopping parameters $t_A$ and $t_B$ are much smaller than the interaction terms, and thus these insulating phases are energetically favored.  
To go beyond the experimental findings, we take the hopping parameters as variables while keeping all interaction terms fixed. 
Although we focus on the one-dimensional case, it is expected that the ordered phases discovered here may persist or even become more stable in two dimensions due to the weakening of the quantum fluctuations.

To obtain the ground-state wave function and the relevant observables,
we employ the DMRG method with open boundary conditions (OBCs) implemented in the ITensor library~\cite{Fishman2022}.
Since the onsite interactions $U_A$ and $U_B$ are much larger than other parameters, we truncate $n_{\text{max}}=2$ for the Fock basis so that $n_{A} (n_{B}) = 0, 1, 2$ at each site.
We have verified that our results are not qualitatively modified when using $n_{\text{max}}=4$.
Unless explicitly mentioned otherwise, we keep the bond dimensions up to $\chi = 2000$, a value that can provide satisfactory precision, with truncation errors on the order of $10^{-10}$ even in gapless phases. The ground-state wave function is ultimately targeted after more than 14 sweeps.

We have calculated the correlation functions and the associated order parameters, the single-particle gap, and the entanglement entropy to map out the quantum phase diagram.
The correlation functions for the SF phase and DW phase are defined as follows
\begin{subequations}\label{EQ:CorrFunc2}
\begin{align}
	&C_{\text{SF}}^{\sigma}(r)=\langle \hat{b}^{\dagger}_{i\sigma}\,\hat{b}_{i+r\sigma}\rangle\\
	&C_{\text{DW}}^{\sigma}(r)=(-1)^{r}\langle\delta\hat{n}_{i\sigma}\,\delta\hat{n}_{i+r\sigma}\rangle \label{CORR:DW}	
 \end{align}
 \end{subequations}
 where $\delta\hat{n}_{i\sigma}=\hat{n}_{i\sigma}-N_\sigma/L$, and $N_{\sigma}$ denotes the number of bosons in component $\sigma$. In the SF phase, $C^{\sigma}_{\text{SF}}(r)$ decays algebraically as $r$, while it is suppressed exponentially in the DW and SCF phases. The DW phase is characterized by a finite $C_{\text{DW}}^{\sigma}(r\to\infty)$. When performing DMRG calculations in the DW phase with OBCs, we apply opposing pinning fields on the boundaries of the Hamiltonian to eliminate the domain walls~\cite{Stumper2020}. The order parameter $O^{\sigma}_{\text{DW}}(L)$ is then obtained by averaging the expected values of all two-point correlation functions $C_{\text{DW}}^{\sigma}(r)$ in the middle part of the chain with a length $L$, discarding $L/4$ sites at both ends. The SCF phase can be distinguished from the PSF phase using correlation functions
\begin{subequations}\label{EQ:CorrFunc4}
\begin{align}
	&C_{\text{SCF}}^{AB}(r)=\langle\hat{b}^{\dagger}_{iA}\,\hat{b}_{iB}\,\hat{b}_{i+rA}\,\hat{b}^{\dagger}_{i+rB}\rangle\\
    &C_{\text{PSF}}^{AB}(r)=\langle\hat{b}^{\dagger}_{iA}\,\hat{b}^{\dagger}_{iB}\,\hat{b}_{i+rA}\,\hat{b}_{i+rB}\rangle.
\end{align}
\end{subequations}
In the SCF~(PSF) phase, $C_{\text{SCF}}^{AB}(r)$ exhibits algebraic (exponential) decay, while $C_{\text{PSF}}^{AB}(r)$ is exponentially (algebraically) suppressed~\cite{Hu2009}.

The single-particle gap $\Delta_{s}^{\sigma}$, defined as the energy difference needed to create a particle excitation and a hole excitation in the system, is commonly used to distinguish the SF phase from insulating phases. For each component, it is given by
\begin{subequations}\label{EQ:SPGap}
\begin{align}
&\Delta_{s}^{A}(L) = \sum_{\upsilon = \pm 1} \big[E_{L}(N_{A} + \upsilon, N_{B}) - E_{L}(N_{A}, N_{ B})\big]\\
&\Delta_{s}^{B}(L) = \sum_{\upsilon = \pm 1} \big[E_{L}(N_{A}, N_{B} + \upsilon) - E_{L}(N_{A}, N_{ B})\big]
\end{align}
\end{subequations}
where $E_L(N_{A},N_{B})$ stands for the ground-state energy of a system with $N_A$ and $N_B$ bosons.
Near the BKT transition, the single-particle gap scales as
\begin{equation}\label{EQ:GAPBKT}
\Delta_{s}^{\sigma}(\infty) \thicksim \exp\bigg(\frac{\text{const}}{\sqrt{|t_{\sigma}^{c}-t_{\sigma}|}}\bigg).
\end{equation}

The gapless feature can also be captured by the von Neumann entanglement entropy $\mathcal{S}\equiv-\text{Tr}(\rho_{\mathcal{A}}\text{ln}\rho_{\mathcal{A}})$, which is readily available in DMRG calculations.
Here, $\rho_{\mathcal{A}}= \text{Tr}_{\mathcal{B}}|\Psi_{\mathcal{AB}}\rangle\langle\Psi_{\mathcal{AB}}|$ is the reduced density matrix of block $\mathcal{A}$
in a bipartite system consisting of system $\mathcal{A}$ and environment $\mathcal{B}$.
The scaling formula of the entanglement entropy $\mathcal{S}$ depends not only on the partition but also on the boundary conditions~\cite{Affleck1991,Tagliacozzo2008,Laflorencie2016}. Notably, it is revealed that OBCs can generate an alternating term in the entanglement entropy which decays away from the boundary algebraically~\cite{Wang2004,Laflorencie2006}. 
Therefore, once a finite chain of total length $L$
is divided into lengths $l$ and $L-l$, it is inferred that~\cite{Luo2018}
\begin{equation}\label{EQ:VNE}
\mathcal{S}_{L}(l)=\frac{c}{6}\,{\text{ln}}\bigg[\frac{2L}{\pi}\,{\text{sin}}\bigg(\frac{\pi l}{L}\bigg)\bigg]+ g^{\prime}D_{l,l+1} + s_{1}^\prime/2,
\end{equation}
where $c$ is the central charge. 
In addition, $g^{\prime}$ and $s_{1}^\prime$ are nonuniversal terms arising from the boundary effect and other correlations~\cite{Affleck1991}, and $D_{l,l+1}$ accounts for the alternating term. 

However, fitting the entanglement entropy $\mathcal{S}_{L}(l)$ via Eq.~\eqref{EQ:VNE} becomes challenging when domain walls exist in certain parameter regions. To eliminate them, we apply pinning fields at the boundaries, rendering boundary effects negligible, and estimate the central charge using the following relation~\cite{Lauchli2008},
\begin{equation}\label{EQ:DiffVNE}
\Delta \mathcal{S}(L) = \mathcal{S}_{L}(L/2)-\mathcal{S}_{L/2}(L/4) = \frac{c}{6}\ln2.
\end{equation}
In contrast, if boundary effects are significant but no domain walls exist, we can treat $D_{l,l+1}=\langle\hat{n}_{l\sigma}\hat{n}_{l+1\sigma}\rangle$ as the correlation function, which is independent of the component $\sigma$, since it shares a similar oscillatory behavior with the entanglement entropy.

\section{NUMERICAL RESULTS}\label{sec:num}

\subsection{Phase transitions at half filling}\label{SEC2:HalfFilling}

\subsubsection{Ground-state phase diagram}\label{SEC3:PhaseDiag}

In this section we fix $\rho_A = \rho_B = 1/2$, where the density of each component is defined as $\rho_{\sigma} = N_{\sigma}/L$. 
In a single-component EBHM with one-half density, it is shown that a strong nearest-neighbor interaction can stabilize a DW phase, in which the lattice translation symmetry is spontaneously broken \cite{Kuhner2000}.
Thus, the DW phase becomes favorable when the hopping parameters $t_A$ and $t_B$ are small concurrently and we abbreviate it as the DW+DW phase. 
On the other hand, when $t_A$ and/or $t_B$ are large enough, bosons prefer staying in SF states. 
It is curious to know how the ground state evolves when tuning $t_A$ and $t_B$.
For this purpose, we fix the interaction terms $U$ and $V$ as those given in Eq.~\eqref{EQ:UabVab} and take $t_\sigma$ as free parameters.
Experimentally, the latter can be tuned by the intensity of the trapping light.
The ground-state phase diagram is plotted in the $t_A$-$t_B$ plane, as is shown in Fig.~\ref{Fig:PhaseDiagram}.

There are five phases in the phase diagram, which includes a DW+DW phase when both $t_A$ and $t_B$ are small, SF+DW and DW+SF phases when one of them becomes large, an SF+SF phase when both are large, and an SCF phase at modest $t_A$ and $t_B$.
The SF+DW, DW+SF, and SCF phases are characterized by a central charge of $c=1$, while it is $c=2$ in the SF+SF phase.
When $t_A$ and $t_B$ are small, the interaction terms are dominant and both types of bosons are then in the DW phase, resulting in a so-called DW+DW phase.
As one of the two hopping parameters, for example, $t_A$, increases,
$A$ bosons enter the SF phase at a critical value $t^c_A$ while $B$ bosons stay in the DW phase,
giving rise to an SF+DW phase.
This transition from the DW+DW phase to the SF+DW phase is analogous to the DW--SF transition, which is of the BKT type,
in the single-component EBHM \cite{Kuhner1998}.
Therefore, the DW+DW--SF+DW transition, DW+DW--DW+SF transition, DW+SF--SF+SF transition, and SF+DW--SF+SF transition in the phase diagram all belong to the BKT universality class.
On the other hand, when $t_A$ and $t_B$ are kept comparable and increase simultaneously,
we can observe two phase transitions. The first one is from the DW+DW to SCF, and the second one is from the SCF to SF+SF phase.
One may notice that in the DW+DW--SF+DW transition, $A$ and $B$ bosons behave almost independently, signifying the absence of intercomponent particle pairing or anti-pairing. However, in the DW+DW--SCF transition the states of both bosons change simultaneously. This occurs because, in the SCF phase $A$ bosons and $B$ holes form a bound state, behaving as a single quasiparticle. The transitions DW+DW--SCF and SCF--SF+SF belong to the BKT universality class. In what follows, we will first disclose properties of SCF and then delve into details of three selected phase transitions pertaining to DW+DW phase, SF+DW phase, SCF phase, and SF+SF phase.

\subsubsection{Supercounterfluid}\label{SEC3:SCF}

\begin{figure}[!t]
\centering
\includegraphics[width=0.95\linewidth]{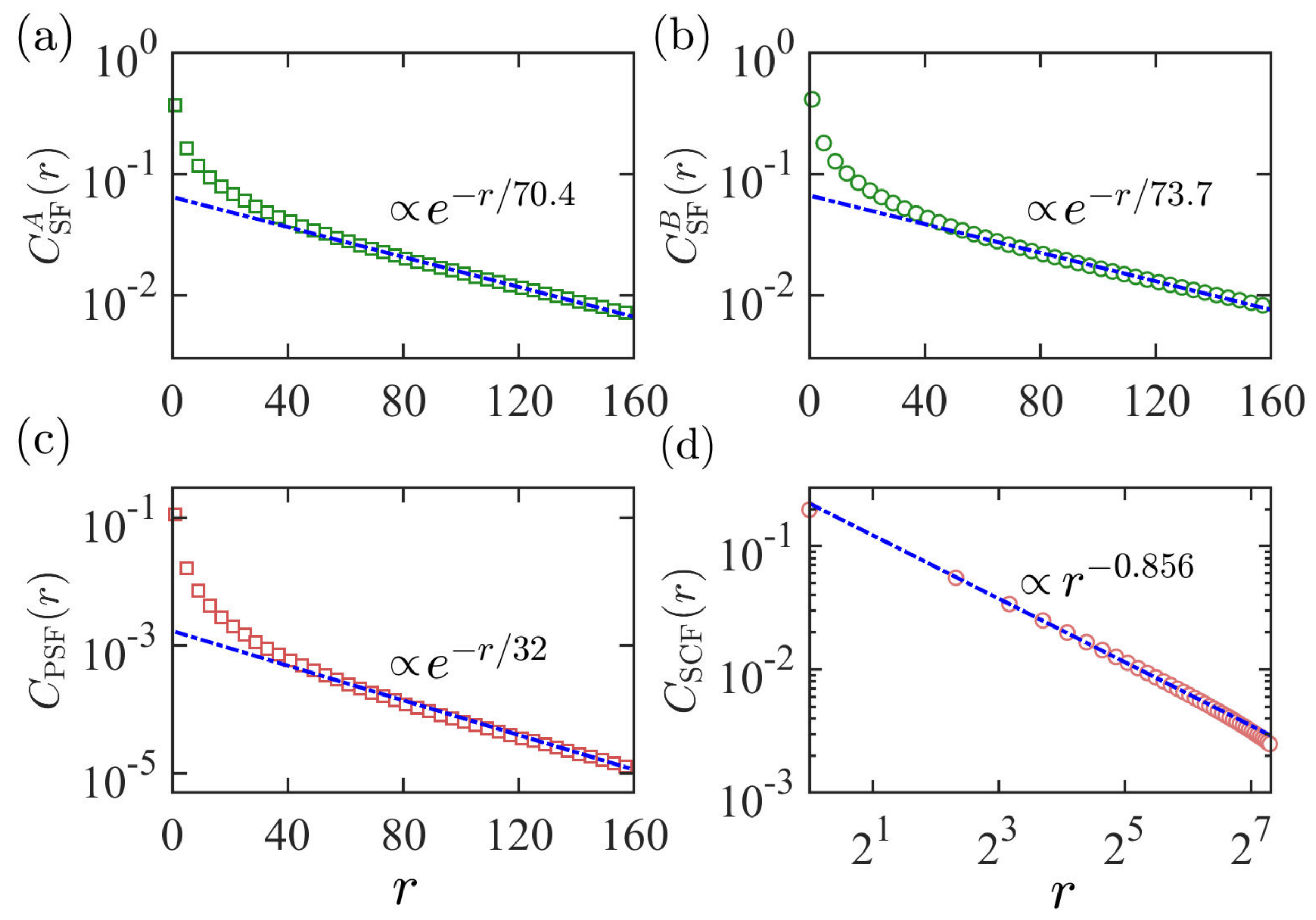}
\caption{Fitting results for the correlation functions $C_{\text{SF}}^{A}(r)$ (a), $C_{\text{SF}}^{B}(r)$ (b), $C_{\text{PSF}}(r)$ (c), and $C_{\text{SCF}}(r)$ (d) in the SCF phase. We choose $t_{A}=0.08$, $t_B=0.12$, and set the truncated bond dimension $\chi=4000$ for $L=320$.}
\label{Fig:CORFUN-SCF}
\end{figure}

In contrast to the trivial Mott insulator with only one species of boson, Mott insulator equipped with two different bosons can form a novel SCF which has nondissipative counterflows in its two components and remains charge neutral in its entirety.
The SCF was initially proposed by Kuklov and Svistunov over two decades ago in the context of a two-component Bose-Hubbard model~\cite{Kuklov2003},
and has incited immense research efforts both theoretically and experimentally~\cite{Kuklov2004b,Hu2011,Ohgoe2011,Sellin2018,Lin2020,Kuklov2004c,Zheng2025}. It is recognized that the SCF is characterized by both the one-body order parameters $\langle\hat{b}_{iA}\rangle=\langle\hat{b}_{iB}\rangle=0$, the pair order $\langle \hat{b}_{iA}\hat{b}_{iB}\rangle=0$, and the anti-pair order $\langle\hat{b}_{iA}^\dagger\hat{b}_{iB}\rangle\neq 0$~\cite{Kuklov2004b}. However, the low-energy behavior of SCF in one dimension resembles that of a Luttinger liquid. We turn to the computation of the correlation functions as defined in Eqs.~\eqref{EQ:CorrFunc2} and~\eqref{EQ:CorrFunc4} to establish this connection. In Figs.~\ref{Fig:CORFUN-SCF}(a) and (b), both $A$ and $B$ bosons decay exponentially in $C_{\text{SF}}^{\sigma}(r)$, but with a relatively large correlation length $\xi^{\sigma}$. The similar behavior of $C_{\text{PSF}}(r)$ is shown in Fig.~\ref{Fig:CORFUN-SCF}(c). Given that $\Delta\sim 1/\xi$, there is a nonvanishing excitation gap associated with particle excitation modes. Furthermore, the algebraic decay of $C_{\text{SCF}}(r)$ shown in Fig.~\ref{Fig:CORFUN-SCF}(d) serves as a hallmark of the existence of the SCF phase.

\begin{figure*}[!t]
\centering	
\includegraphics[width=0.95\linewidth]{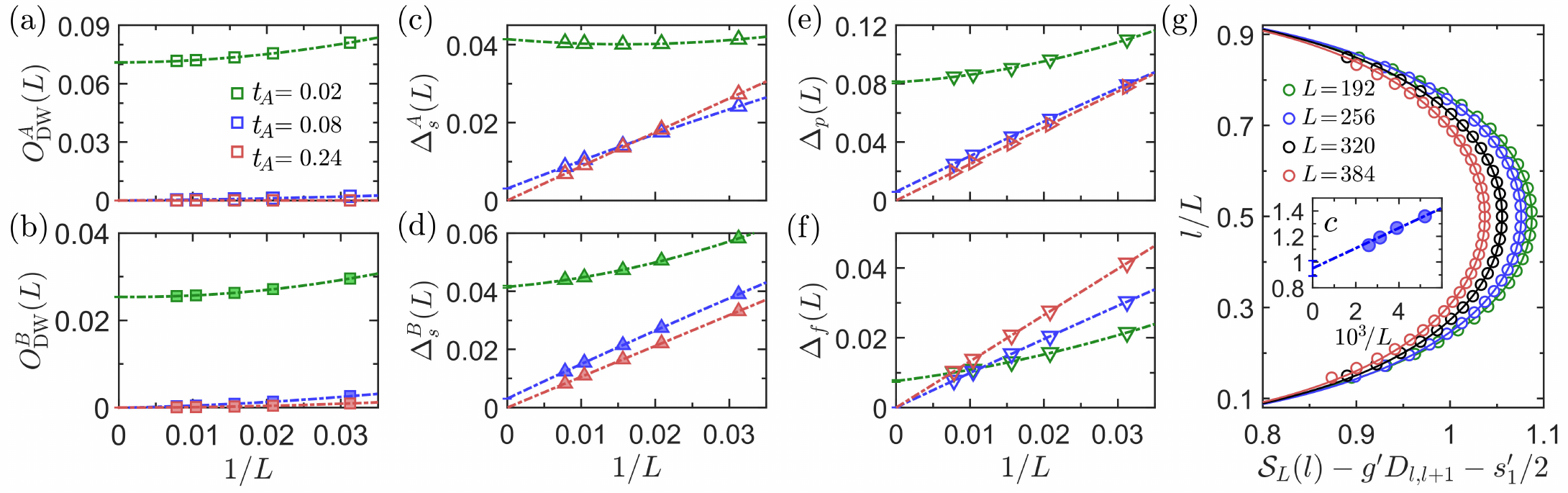}
\caption{(a)-(f) The size dependence of the order parameter $O_{\text{DW}}^{\sigma}$, the single-particle gap $\Delta_{s}^{\sigma}$, the two-particle gap $\Delta_{p}$, and the particle-flipping gap $\Delta_{f}$ are extrapolated using a quadratic polynomial form to the large $L$ limit  in three phases: DW+DW, SCF and SF+SF.  These correspond to $t_A=0.02,0.08,0.24$, respectively, with $t_B$ fixed at $=0.12$. 
(g) In the SCF phase, the central charge is extracted by fitting entanglement entropy $\mathcal{S}_{L}(l)$ using Eq.~\eqref{EQ:VNE}. Inset of (g) shows the linear extrapolation to the large $L$ limit for finite system sizes $L=192,256,320,384$, resulting $c=0.95(6)$. The truncated bond dimension is set to $\chi=4000$ for the entanglement entropy calculation.}
\label{Fig:SCF}
\end{figure*}

Previous theoretical works proposed that the DW pattern can either coexist with SCF or be absent in the two-component Bose-Hubbard model. The former corresponds to a supersolid of antipairs, i.e., the SCF+DW phase, while the latter corresponds to the SCF phase~\cite{Mathey2007,Mathey2009,Hu2009,Trefzger2009}.
To gain deeper insight into the SCF phase identified in the present work, we compare its properties with those of the other two phases, DW+DW and SF+SF, along the cut line $t_B=0.12$ in Fig.~\ref{Fig:PhaseDiagram} by examining the order parameters, excitations, and central charge. 
We first calculate the order parameters $O_{\text{DW}}^{A/B}$ of DW ordering at three selected parameters $t_A$ = 0.02, 0.08, and 0.24 in Figs.~\ref{Fig:SCF}(a) and (b).
The calculations of single-particle gaps $\Delta_{s}^{A/B}$ with the same parameters are shown in Figs.~\ref{Fig:SCF}(c) and (d).
The fact that both $O_{\text{DW}}^{A/B}$ and $\Delta_{s}^{A/B}$ are finite (zero) is consistent with the underlying DW+DW (SF+SF) phase at $t_A$ = 0.02 (0.24). 
By contrast, in the intermediate phase at $t_A$ = 0.08, $O_{\text{DW}}^{A/B}$ are zero while $\Delta_{s}^{A/B}$ are finite but small due to the weak hopping parameters. Therefore, the possible coexistence with DW pattern is excluded. To further unveil the excitations of the intermediate phase, we resort to the excitation gaps beyond single particle. 
One is the particle-flipping gap $\Delta_f(L)=\Delta_{f}^{A\to B}(L)+\Delta_{f}^{B\to A}(L)$ in which
\begin{subequations}\label{EQ:SPGap}
\begin{align}
&\Delta_{f}^{A\to B}(L)=E_{L}(N_{A}-1,N_{B}+1)\!-\!E_{L}(N_{A},N_{B}),\\
&\Delta_{f}^{B\to A}(L)=E_{L}(N_{A}+1,N_{B}-1)\!-\!E_{L}(N_{A},N_{B}).
\end{align}
\end{subequations}
The other is the two-particle gap, defined by the creation or annihilation of a particle pair from the ground state,
\begin{align}
\Delta_{p}(L) = \sum_{\upsilon = \pm 1} \big[E_{L}(N_{A}+\upsilon, N_{B}+\upsilon) - E_{L}(N_{A}, N_{ B})\big]  
\end{align}
It is observed in Figs.~\ref{Fig:SCF}(e) and (f) that the intermediate phase at $t_A$ = 0.08 yet has a finite two-particle gap $\Delta_{p}$ but a vanishing particle-flipping gap $\Delta_f$. 
The closure of the particle flipping gap corresponds exactly to the existence of quasi-long-range order in $C_{\text{PSF}}(r)$, as observed in Fig.~\ref{Fig:CORFUN-SCF}(d), indicating that the intermediate phase is the sought-after SCF. 
Finally, we extract the value of central charge in the SCF phase.
In light of the artful trick which eliminates the oscillations in entanglement entropy by leveraging correlation functions $D_{l,l+1}$ in Eq.~\eqref{EQ:VNE}, one can yield  well-shaped curves from which the central charge can be estimated.
As can be seen from Fig.~\ref{Fig:SCF}(g), the linear extrapolation of these finite-size central charges on a series of chain lengths $L$ ranging from 192 to 384 yields a value of $c=0.95(6)$, which agrees with the expected value $c=1$ since there is one gapless mode associated with the particle-flipping excitation.

\subsubsection{DW+DW--SF+DW transition}\label{SEC3:DW2SF}

\begin{figure}[!h]
\centering	
\includegraphics[width=0.98\columnwidth]{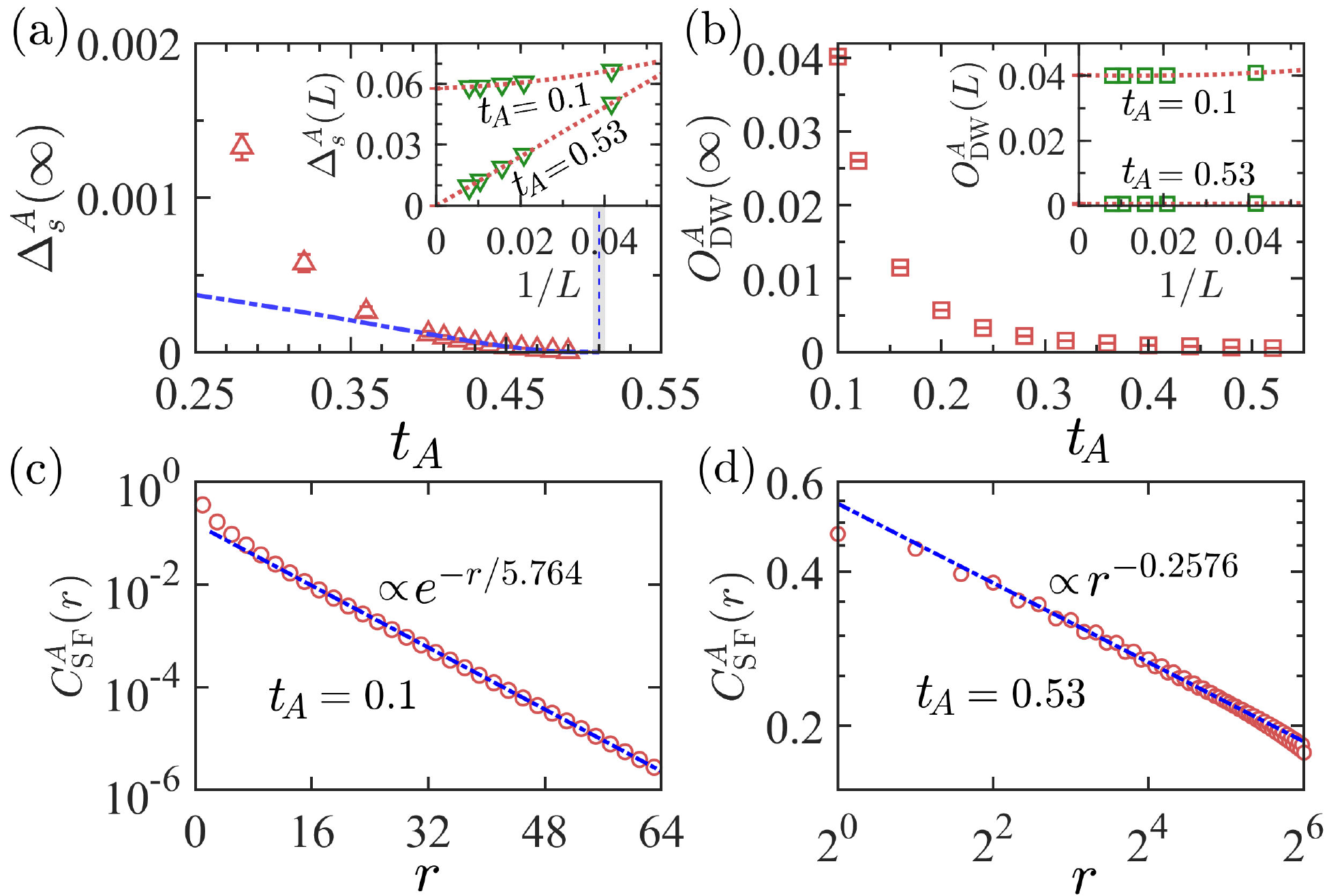}
\caption{(a)-(b) $\Delta_{s}^{A}(\infty)$ and $O_{\text{DW}}^A(\infty)$ are shown as functions of $t_A$, with $t_B$ fixed at $0.04$. The inset shows values extrapolated to the large $L$ limit for $t_{A}=0.1$ and $t_{A}=0.53$. The blue dash-dot line in (a) represents the best fit of $\Delta_{s}^{A}(\infty)$ near the DW+DW--SF+DW phase transition and follows the formula $\Delta_{s}^{A}(\infty,t_{A})\!=\!0.003\!\times\exp\big(\!-\!1.091/\sqrt{0.510-t_{A}}\,\big)$. (c) and (d) show the decay of $C^{A}_{\text{SF}}(r)$ with $r$ at $t_{A} = 0.1$ and $t_{A} = 0.53$, respectively.}
\label{Fig:DW2SF}
\end{figure}

We carry on elucidating the nature of the DW+DW--SF+DW transition on the cut line at $t_B=0.04$, as shown in Fig.~\ref{Fig:PhaseDiagram}. It is used as an example that involves only single-component bosons transitioning from DW to SF, while the other component remains unchanged.
Since the hopping parameter of $B$ bosons are rather small, we expect that there is not a mixture of the two species of bosons when tuning hoping parameter $t_A$. 
Thus, this transition can be captured by considering physical properties of $A$ bosons only.
We plot the single-particle gap $\Delta_c^{A}(\infty)$ as a function of $t_A$ in Fig.~\ref{Fig:DW2SF}(a). 
At each $t_A$, $\Delta_c^{A}(\infty)$ is obtained by a proper extrapolation of $\Delta_c^{A}(L)$ at different chain lengths $L$, as shown in the inset for $t_A$ = 0.1 and 0.53.
As $t_A$ increases, $\Delta_c^A(\infty)$ undergoes a  monotonous decrease whose value is finite in the DW+DW phase and is zero in the SF+DW phase.
In particular,  $\Delta_c^A(\infty)$ closes at a critical value $t_A^c$ where the DW+DW--SF+DW transition occurs, 
and near the critical point the gap can be well fitted by $\Delta_{s}^{A}(\infty,t_{A}) = 0.003 \times \exp\big(-1.091/\sqrt{t_A^c - t_{A}}\big)$ with $t_A^c=0.510(4)$.
Such a behavior of the excitation gap indicates that this transition should be of BKT type. 
To further confirm the nature of these two phases, one can calculate the DW order parameter $O_{\text{DW}}^\sigma$ and SF correlation function $C_{\text{SF}}^\sigma$. 
For this purpose, we show the $A$-type DW order parameter $O_{\text{DW}}^A(\infty)$ as a function of $t_A$ in Fig.~\ref{Fig:DW2SF}(b). 
Likewise, two representative points $t_A = 0.10$ and $t_ A= 0.53$ are chosen again with one smaller than $t_A^c$ and the other larger than $t_A^c$, respectively.
As shown in the inset, $O_{\text{DW}}^A(\infty)$ is finite at $t_A=0.10$ while it vanishes at $t_A=0.53$,
indicating that $A$ bosons are in a DW phase only at $t_A=0.10$.
On the other hand, at both points $O_{\text{DW}}^B(\infty)$ is finite (not shown), suggesting that $B$ bosons are always in the DW phase.
Figs.~\ref{Fig:DW2SF}(c) and (d) show the $A$-type SF correlation function $C_{\text{SF}}^A$ at $t_A$ = 0.10 and 0.53, respectively. 
The $C_{\text{SF}}^A(r)$ decays exponentially at $t_A$ = 0.10 while it decays algebraically at $t_A = 0.53$.
We then conclude that in the latter case $A$ bosons are in the SF phase.

\begin{figure}[!t]
	\centering	
	\includegraphics[width=0.90\columnwidth]{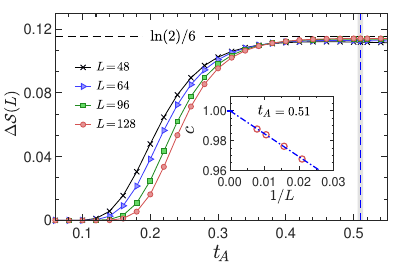}
	\caption{$\Delta{S}_L$ is shown as a function of $t_A$ for $t_{B}=0.04$, with $L=48,\,64,\,96,$ and $128$. As $t_A$ increases $\Delta\mathcal{S}(L)$ approaches $\ln{(2)}/6$, indicating the SF+DW phase. In the thermodynamic limit, we obtain  $t_{A}^{c}=0.510(4)$. The inset shows the central charge $c=1$ at $t_{A}=0.510$.}
	\label{Fig:DeltaVNE}
\end{figure}

The occurrence of DW+DW--SF+DW transition can also be confirmed by computing $\Delta \mathcal{S}(L)$ defined in Eq.~\eqref{EQ:DiffVNE}. As we know, the entanglement entropy $\mathcal{S}_L$ saturates in a gapped system when the block size $L$ is much
larger than the correlation length $\xi$, and one obtains $\Delta \mathcal{S}(L) \simeq 0$. 
However, in a critical SF phase it scales logarithmically and one expects $\Delta \mathcal{S}(L)=\frac{c}{6} {\text{ln}}\,2$ with $c=1$ in the large $L$ limit.
Fig.~\ref{Fig:DeltaVNE} illustrates the behavior of $\Delta \mathcal{S}(L)$ as a function of the hopping parameter $t_A$.
$\Delta \mathcal{S}(L)$ scales to zero with increasing system length $L$ in the DW+DW phase, while it converges towards the expected value of $\Delta \mathcal{S}(L\to\infty) = \ln(2)/6$ in the SF+DW phase. 
In the inset of Fig.~\ref{Fig:DeltaVNE}, we obtain the central charge $c\simeq{1.0}$ at $t_A^c = 0.510(4)$, conforming to the conclusion that only one component of boson is SF.

\subsubsection{DW+DW--SCF transition}\label{SEC3:DW2SCF}

\begin{figure}[!t]
	\centering	
	\includegraphics[width=0.98\columnwidth]{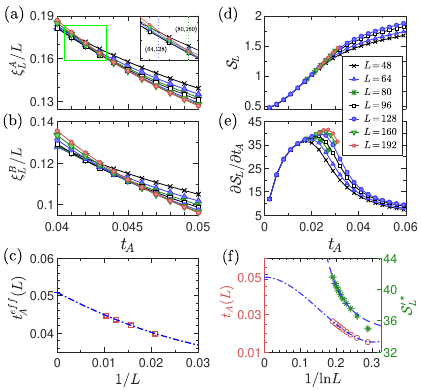}
	\caption{(a)-(b) Finite size scaling of $\xi_{L}^{\sigma}/L$ is plotted as a function of $t_A$, with $t_B$ fixed at $0.12$. The inset shows a zoomed-in view of the crosspoints in a green rectangular region. (c) The DW+DW--SCF transition point is found at $t_A=0.05089(2)$. (d)-(e) $\mathcal{S}_L$ and $\partial{\mathcal{S}_L}/\partial{t_A}$ are shown as functions of $t_{A}$. (f) The peak locations and heights of $\partial{\mathcal{S}_L}/\partial{t_A}$ are extracted from (e) at $t_{A}(L)$. Extrapolation of the former to the large $L$ limit yields $t_A^{c}=0.0499(16)$, according to Eq.~\eqref{EQ:DerVNEInfOrder}, and the latter is fitted via Eq.~\eqref{EQ:DerVNEHeight}. We set the truncated bond dimension $\chi=3000$.}
	\label{Fig:DW2SCF}
\end{figure}

\begin{figure*}[!t]
\centering	
\includegraphics[width=1.7\columnwidth]{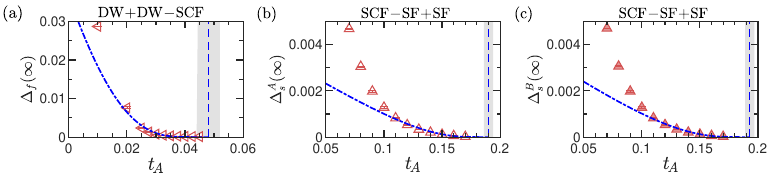}
\caption{The closing behavior of the energy gaps near the DW+DW--SCF and SCF--SF+SF phase transition is shown in (a) $\Delta_{f}(\infty)$, (b) $\Delta_{s}^{A}(\infty)$ and (c) $\Delta_{s}^{B}(\infty)$, with the hopping parameter $t_{B}$ fixed at $0.12$. The blue dash-dot line represents the best fit for  $\Delta_{f}(\infty,t_{A})=9.999\times\!\exp\big(\!-1.229/\sqrt{0.048-t_{A}}\big)$,  $\Delta_{s}^{A}(\infty,t_{A})=0.068\times\!\exp\big(\!-1.263/\sqrt{0.190-t_{A}}\big)$, and $\Delta_{s}^{B}(\infty,t_{A})=0.086\times\!\exp\big(\!-1.354/\sqrt{0.193-t_{A}}\big)$.}
\label{Fig:SCFGapFit}
\end{figure*}

We proceed to study the DW+DW--SCF transition by focusing on the cut line at $t_B=0.12$.
Given that natures of the underlying phases have already been unraveled in earlier paragraphs, our aims at present are to determine the transition point and to possibly identify the universality class.
To commence with, since the DW+DW phase is gapped and the SCF is gapless, a continuous transition between them is possible. We can leverage correlation lengths of both $A$ and $B$ bosons to discern the two. For a sufficiently large system, the second-moment correlation length at a small momentum point $k$ is estimated by \cite{Cuccoli1995},
\begin{equation}
\xi_{L}^{\sigma}=\frac{1}{\delta k}\sqrt{\frac{S_{k}^{\sigma}(L)}{S_{k+\delta k}^{\sigma}(L)}-1},
\end{equation}
where  
\begin{equation}\label{Eq:SSF}
        S_{k}^{\sigma}(L)=\frac{1}{L^{2}}\sum_{i,j} e^{-ik(i-j)}\langle\delta\hat{n}_{i\sigma}\delta\hat{n}_{j\sigma}\rangle
\end{equation}
is the structure factor and $\delta k=2\pi/L$. We choose $k = \pi$, which dominates the structure factor peak in the DW+DW phase, so that $L^2S_{\pi}^{\sigma}(L)$ equals the sum of all two-point correlations defined in Eq.~\eqref{CORR:DW}. At any critical point, the correlation length diverges. Therefore, the finite-size scaling of $\xi_{L}^{\sigma}/L$ becomes independent of system size near the critical point $t_{A}^{c}$, allowing $t_{A}^{c}$ to be determined in an unbiased manner. The scaled quantities $\xi_{L}^{A}/L$ 
and $\xi_{L}^{B}/L$ as functions of $t_{A}$ are shown in Figs.~\ref{Fig:DW2SCF}(a) and (b), respectively.
The chain lengths chosen are $L$ = 48, 64, 80, 96, 128, 160, 192. In each panel, the curves $\xi_{L}^{\sigma}/L$ exhibit multiple crossings and the crosspoints have a tendency to converge with the increase of $L$. Nevertheless, these crosspoints for the two bosons are different in the finite-size cases.
To reconcile the two, we focus on the effective correlation length $\xi_{L}^{eff} = \sqrt{[(\xi_L^A)^2 + (\xi_L^B)^2]/2}$ and take the crosspoint between $\xi_{L}^{eff}/L$ and $\xi_{2L}^{eff}/(2L)$ denoted by $t_A^{eff}(L)$ as a reference point.
Figure~\ref{Fig:DW2SCF}(c) shows four crosspoints $t_A^{eff}(L)$ with $L$ = 48, 64, 80, and 96.
The quadratic polynomial fitting yields an estimated critical point $t_A = 0.05089(2)$.

Meanwhile, the entanglement entropy is also a versatile tool to determine the critical point. 
Figure~\ref{Fig:DW2SCF}(d) presents the entanglement entropy $\mathcal{S}_L$ as a function of $t_A$ for a series of chain lengths $L$.
The $\mathcal{S}_L$ grows gradually from the gapped DW+DW phase to the gapless SCF phase. It is almost size-independent deep into the DW+DW phase while it initiates bifurcation when entering into the SCF phase.  Instead of using entanglement entropy directly, we hereby exploit its derivative $\partial{\mathcal{S}_L}/\partial{t_A}$ to pin down the critical point. 
As shown in Fig.~\ref{Fig:DW2SCF}(e), $\partial{\mathcal{S}_L}/\partial{t_A}$ increases monotonically and reaches a peak, after which it decreases monotonically. 
Consequently, the critical point can be inferred from the divergent point in $\partial{\mathcal{S}_L}/\partial{t_A}$.
The derivative of entanglement entropy has been employed to systematically investigate infinite-order phase transitions in the quantum $O(2)$ model, achieving some success in locating the critical point and identifying its universality class for different values of spin-$S$~\cite{Zhang2021}. Along this line, we find that the scaling of the peak position $t_{A}(L)$ of $\partial{\mathcal{S}_L}/\partial{t_A}$ can be well fitted by including a correction term proportional to $1/\ln^3(L)$ for the BKT transition, 
\begin{equation}\label{EQ:DerVNEInfOrder}
 t_A^c = t_{A}(L) + \frac{b_s^2}{\ln^2(L)} + \frac{d_s}{\ln^3(L)} + \cdots,
\end{equation}
where $b_s$ and $d_s$ are constants. Fig.~\ref{Fig:DW2SCF}(f) shows $t_A(L)$ as a function of $1/\ln{L}$ and extrapolates it to the large $L$ limit via Eq.~\eqref{EQ:DerVNEInfOrder}, resulting in $t_A^c=0.0499(16)$. 
The peak height of entanglement entropy’s derivative ${\mathcal{S}_{L}^{\prime}}^{\!*}$ is readily obtained by computing partial derivatives at $t_{A}(L)$ on both ends of Eq.~\eqref{EQ:VNE},
\begin{align}\label{EQ:DerVNEHeight}
{\mathcal{S}_{L}^{\prime}}^{\!*}&=\frac{c}{6}\frac{\partial{L}}{L\partial{t_{A}(L)}}+r^{\prime}=\frac{a_{s}\ln^3(L)}{1+d_{s}^{\prime}/\ln(L)+\cdots}+r_{s}^{\prime},
\end{align}
where $a_s=c/(12b_{s}^{2})$, $d_{s}^{\prime}$ and $r_{s}^{\prime}$ are constants. 
We see that ${\mathcal{S}_{L}^{\prime}}^{\!*}$ plotted as a function of $1/\ln{L}$, agrees well with the scaling form of Eq.~\eqref{EQ:DerVNEHeight}. Except for $L\leq 48$, nearly all data points reside along the fit line. After carrying out Eqs.~\eqref{EQ:DerVNEInfOrder} and \eqref{EQ:DerVNEHeight}, we extract the values of $b_{s}$ and $a_{s}$, and obtain $c\simeq0.99$ by exploiting $a_s=c/(12b_{s}^{2})$.

In addition, the critical point can also be obtained via the excitation gap. In Fig.~\ref{Fig:SCFGapFit}(a) we plot the particle-flipping gap $\Delta_{f}(\infty)$ as a function of $t_A$. Near the transition point, it can be well fitted by the formula $9.999\times\!\exp\big(\!-1.229/\sqrt{t_A^c-t_{A}}\big)$ with $t_A^c=0.048(4)$. 
This value is closely consistent with the one obtained in Fig.~\ref{Fig:DW2SCF} within the error.
Thus, we conclude that such a phase transition is of BKT type.

\begin{figure}[!b]
	\centering	
	\includegraphics[width=0.90\columnwidth]{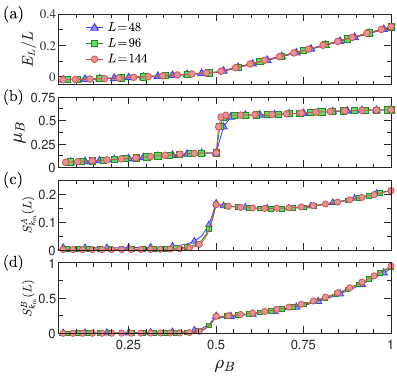}
	\caption{$E_L, \mu_{B}$, $S_{k}^{\sigma}(L)$ are plotted as functions of $\rho_{B}$. The parameters are fixed at $t_A=2t_{B}=0.04$ and $\rho_A=1/2$. (a) The kink in $E_L$ at $\rho_B=1/2$ indicates a first-order phase transition of SF+SF--DW+SS. (b) A sudden jump in $\mu_B$ occurs at $\rho_B=1/2$, suggesting the presence of a DW+DW phase. (c)-(d) $S_{k_m}^{\sigma}$ is finite when $\rho_{B}\ge{1/2}$ and it approaches zero when $\rho_B<1/2$. $k_m$ is the peak position of $S_{k_m}^\sigma(L)$ in the momentum space. Its value is $\pi$ when $\rho_{B}\ge{1/2}$.}
	\label{Fig:RhoBTuned}
\end{figure}

\begin{figure*}[!t]
    \includegraphics[width=0.92\linewidth]{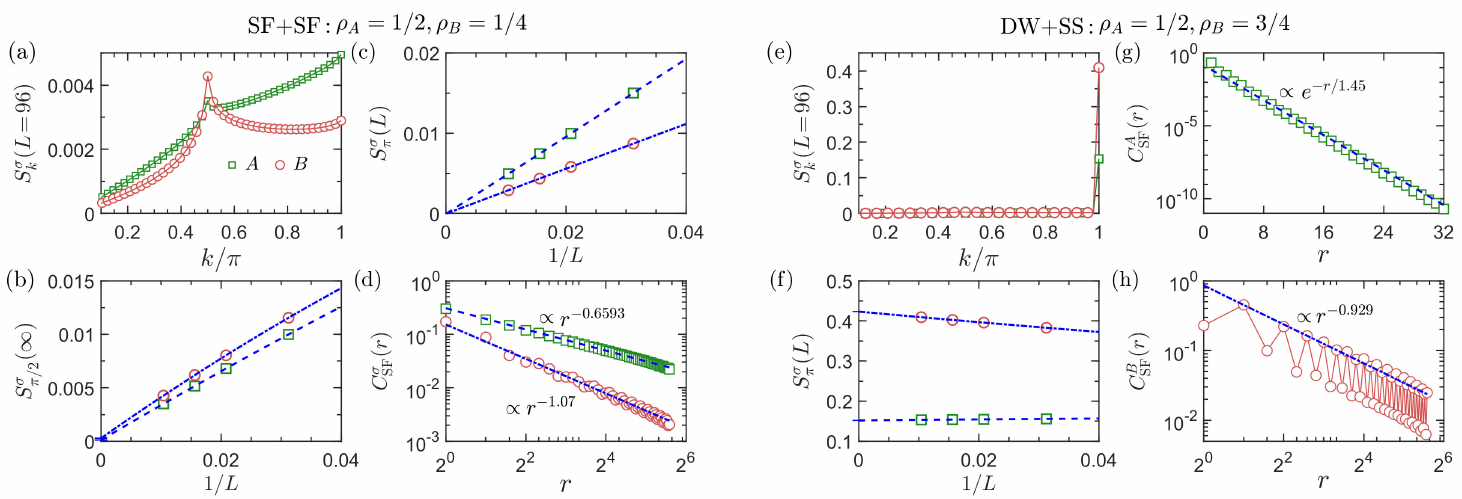}
    \caption{The quantities $S_{k}^{\sigma}$ and $C_{\text{SF}}^{\sigma}(r)$ characterize features of the SF+SF phase in (a)-(d) and the DW+SS phase in (e)-(h), which correspond to $\rho_{B}=1/4$ and $\rho_{B}=3/4$, respectively. The other parameters are fixed at $\rho_{A}=1/2$ and $t_A=2t_{B}=0.04$. (b) and (c) show the peaks of $S_{k}^{\sigma}(L)$ in (a) vanish in the large $L$ limit. (d) shows $C_{\text{SF}}^{A}(r)$ and $C_{\text{SF}}^{B}(r)$ decays exponentially. (f) shows the peak of $S_{k}^{\sigma}(L)$ in (e) is nonzero in the large $L$ limit. (g) shows $C_{\text{SF}}^{A}(r)$ decays exponentially, while $C_{\text{SF}}^{B}(r)$ decays algebraically in (h).}\label{Fig:RhoB1Quadvs3Quad}
\end{figure*}

\subsubsection{SCF--SF+SF transition}\label{SEC3:SCF2SF}

In the end, we study the SCF--SF+SF transition, which is also of BKT type, by calculating the single-particle gap.
The single-particle gaps $\Delta_{s}^{A/B}(\infty)$ as functions of $t_{A}$ at fixed $t_B = 0.12$ are shown in Figs.~\ref{Fig:SCFGapFit}(b) and (c).
By fitting the two single-particle gaps with Eq.~\eqref{EQ:GAPBKT}, it is inferred that the critical points are $t_{A} \simeq 0.190(5)$ and 0.193(5), respectively.
As a result, we can safely conclude that the transition occurs simultaneously for both types of bosons. 
In the gapless SF+SF phase in which both bosons are critical, one expects the central charge to be 2.
The fitting of the central charge from entanglement entropy indeed gives $c = 2$ (not shown).

\subsection{Phase transitions at incommensurate fillings}\label{sec:den}

In this subsection, we investigate the evolution of quantum phases starting from the DW+DW phase with $\rho_{A} = \rho_{B} = 1/2$ as the density of bosons in one component changes.
Previous works have shown that the insulating DW phase can exist only at some commensurate fillings~\cite{Mishra2009,Sengupta2005}.
We thus expect that changing the particle density will drive it into other phases.
For this purpose we fix $\rho_{A}=1/2$ and vary $\rho_{B}$ from $0$ to $1$.
Physically, the value of $\rho_B$ can be controlled by the chemical potential $\mu_B$.
At filling $\rho_B=N_B/L$, $\mu_B$ is given by $\mu_B(\rho_B) \!= E_L(N_A,N_B)-E_L(N_A,N_B\!-1)$.
In addition, unlike the case at commensurate filling, the DW order parameter  $O_{\text{DW}}^{\sigma}$ defined according to Eq.~\eqref{CORR:DW} may fail.
Instead, at a general filling the DW order parameter can be measured by the structure factor defined in Eq.~\eqref{Eq:SSF}.
We then pick the peak heights of $S_{k}^\sigma(L)$ at $k_m\neq 0$ in the momentum space and extrapolate them to the large $L$ limit.
A finite value at $k_m$ suggests the existence of DW ordering.

By setting the hopping parameters $t_A = 2t_{B} = 0.04$, which are comparable to the smallest interaction terms $V_A$ and $V_C$. Since at half filling the system is in a DW+DW state (cf. Fig.~\ref{Fig:PhaseDiagram}), we start from this to calculate the ground-state energy per site $E_L/L$, the chemical potential $\mu_B$, and the structure factor $S_{k}^{\sigma}$ as functions of $\rho_B$ in Fig.~\ref{Fig:RhoBTuned}.
It is observed from Fig.~\ref{Fig:RhoBTuned}(a) that there is a kink in the energy curve at $\rho_B = 1/2$, showing that a first-order phase transition occurs.
The plot of $\mu_{B}$ vs $\rho_{B}$ in Fig.~\ref{Fig:RhoBTuned}(b) shows that $\mu_B$ grows monotonically as $\rho_B$ increases.
In particular, a jump at $\rho_{B} = 1/2$ is visible, which again signals a first-order phase transition. Such a discontinuity tells us that
at $\rho_B = 1/2$ the ground state is an insulator.
Apart from the special point at $\rho_B = 1/2$, $\mu_B$ depends on $\rho_B$ continuously at other fillings, indicating that $B$ bosons always have a SF behavior.
The structure factors $S_{k_m}^{A}(L)$ and $S_{k_m}^{B}(L)$ shown in Fig.~\ref{Fig:RhoBTuned}(c) and (d) also exhibit alluring behaviors.
When $\rho_B \ge {1/2}$, the peak of $S_k^\sigma(L)$ appears at $k_m = \pi$,
and DW order parameters of both components are finite.
Nevertheless, $S_{\pi}^{B}(L)$ increases monotonically while $S_{\pi}^{A}(L)$ is approximately constant as $\rho_B$ varies.
By contrast, when $\rho_B < 1/2$ the peak position $k_m$ of $S_k^\sigma(L)$ depends on the filling $\rho_B$. Numerically, it is inferred that $k_m = 2\rho_B\pi$.
Despite the changeability of $k_m$, DW order parameters are vanishingly small for both components.
We conclude that the ground state is in the SF+SF phase when $\rho_B < 1/2$ and in the DW+SS phase when $1/2 < \rho_B < 1$.

To further confirm the nature of the phases at small and large fillings around $\rho_B = 1/2$, we take $\rho_B = 1/4$ and $\rho_B = 3/4$ as typical examples to show some details, see Fig.~\ref{Fig:RhoB1Quadvs3Quad}.
Here, panels (a)-(d) are for the case of $\rho_B = 1/4$, while  panels (e)-(h) are for the case of $\rho_B = 3/4$.
Fig.~\ref{Fig:RhoB1Quadvs3Quad}(a) shows the structure factors $S_{k}^{A}$ and $S_{k}^{B}$.
The $S_{k}^{A}$ peaks at $k = \pi$, with a subleading peak at $k = \pi/2$ due to the distraction by $B$ bosons. 
In line with the periodic distribution of $B$ bosons, $S_{k}^{B}$ has an apparent peak at $k = \pi/2$ besides $k = \pi$. 
However, the unique crystal order that appears in $A$ and $B$ components is quite fragile. 
As illustrated in Figs.~\ref{Fig:RhoB1Quadvs3Quad}(b) and (c), $S_{k}^{A}(L)$ and $S_{k}^{B}(L)$ with $k = \pi/2$ and $k = \pi$ vanish in the large $L$ limit, showing the absence of DW ordering. 
Further, the algebraical decay of $C_{\text{SF}}^\sigma(r)$ shown in Fig.~\ref{Fig:RhoB1Quadvs3Quad}(d) suggests the presence of SF behavior.
Thus, we conclude that both $A$ and $B$ bosons are in the SF+SF phase.
On the other hand, when $\rho_B = 3/4$ there are only one peak at $k=\pi$ in $S_k^\sigma(L)$ for both $A$ and $B$ components, see Fig.~\ref{Fig:RhoB1Quadvs3Quad}(e).
Further, $S_\pi^A(L)$ and $S_\pi^B(L)$ are robust against the system size $L$, and their extrapolations to the large $L$ limit both yield finite values [see Fig.~\ref{Fig:RhoB1Quadvs3Quad}(f)].
This makes sense that both $A$ and $B$ bosons have DW orders.
Yet, the SF correlation function exhibits rather different behaviors for $A$ and $B$ bosons.
As shown in Figs.~\ref{Fig:RhoB1Quadvs3Quad}(g) and (h), $C_{\text{SF}}^\sigma(r)$ possesses an exponential decay for $A$ bosons and an algebraical decay for $B$ bosons, respectively.
This indicates the absence of SF order in $A$ bosons but the existence of SF order in $B$ bosons.
Noteworthy, the coexistence of DW order and SF order in $B$ bosons enforcing the formulation of SS.
It is in this sense that we call the entirety as a DW+SS phase.

\begin{figure}[!t]
    \centering	
    \includegraphics[width=0.90\columnwidth]{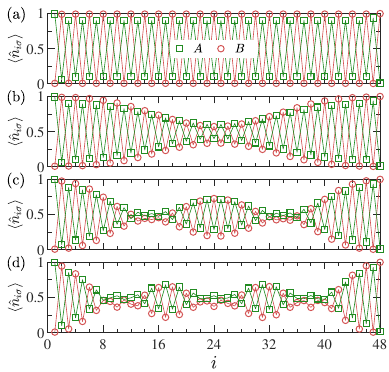}
	\caption{(a) shows the local density $\langle\hat{n}_{i\sigma}\rangle$ as a function of lattice sites $j$ in the DW+DW phase, with $\rho_{A}=\rho_{B}=1/2$ and $t_A=2t_{B}=0.04$. Doping one (b), two (c), or three holes (d) of type $B$ bosons from this state generates one, two, or three pairs of solitons, respectively.}
	\label{Fig:Soliton}
\end{figure}

In the DW+DW phase, $B$ bosons are distributed with the occupation patterns $|\alpha,1\!-\!\alpha,\alpha,1\!-\!\alpha,\cdots\rangle$.
Doping one hole generates a pair of mobile solitons.
When two holes are doped, two pairs of mobile solitons emerge. With further doping, additional soliton pairs appear~\cite{Batrouni2006,Mishra2009}.
These soliton pairs destroy the periodic DW pattern and $B$ bosons are driven into SF phase. Although in the doping process the density of $A$ bosons $\rho_A=1/2$ is fixed,
it is also driven into the SF phase. This is because the occupation of $A$ should change correspondingly to minimize the energy when the pattern of $B$ bosons changes.
The process of forming soliton pairs is shown in Fig.~\ref{Fig:Soliton}. We can see that the relation $\langle n_{iB}\rangle<1/2$ when $\langle n_{iA}\rangle >1/2$ and vice versa are roughly satisfied. This is because the intercomponent interactions $U_{AB}$ and $V_{AB}$ are positive.

\section{\label{sec:con}conclusions}

In this work, we have investigated ground-state properties of a two-component EBHM with strong nearest-neighbor dipolar interactions at different filling factors by analyzing the long-distance behavior of correlation functions, scaling formulas of excitation gaps, and size effect of entanglement entropy.
On the one hand, when both boson components are at half filling,
we identify five distinct phases, which are a DW+DW phase, an SCF phase, a DW+SF phase, an SF+DW phase, and an SF+SF phase, by adjusting the hopping parameters.
Regarding the rich phase diagram, one of its hallmarks is that it indeed attains a DW+DW ground state at weak hopping parameters as revealed in a relevant cold atom experiment albeit on two-dimensional optical lattice.
Another fascinating finding is that it hosts a sought-after SCF, which is also  realized experimentally via a binary Bose mixture in optical lattices.
Pertaining to the phase transitions, we manage to identify the 
DW+DW--SF+DW transition and SCF--SF+SF transition both belong to the BKT universality class, yet with a central charge of 1 and 2, respectively.
Further, critical point of the DW+DW--SCF transition is determined consistently by the correlation lengths and entanglement entropy.
On the other hand, we also discuss the case in which one of the boson densities, say $B$ component, deviates from half filling. 
There is an SF+SF phase when $\rho_{B} < 1/2$ and a DW+SS phase when $1/2 < \rho_{B} < 1$.
Remarkably, the SS phase, which is characterized by the coexistence of diagonal and off-diagonal orders, is realized when tuning the particle density in a wide range.
In our study, this phase is confirmed by analyzing the staggered structure factor $S_{\pi}^{\sigma}(L)$ in the large $L$ limit and examining the decay behavior of the SF correlation function $C_{\text{SF}}^{\sigma}(r)$.

So far, it is firmly believed that these systems with ultracold atoms in optical lattices serve as ideal quantum simulators to realize exotic phases of matter.
In addition, the EBHM is particularly prevailing contemporarily in the interpretation of ultracold atom experiments.
Therefore, our study of the rich phase diagram of the two-component EBHM provides clues of hunting for interesting phases exemplified by the SCF phase.
Although there are some proposals of creating SCF, most of them require unit fillings of boson components. 
Our work thus paves the way for producing SCF with fractional filling factors.

\begin{acknowledgments}
This work is supported by the National Key R\&D Program of China (Grants No. 2022YFA1402704),
the National Natural Science Foundation of China (Grants No. 12304176, No. 12274187, and No. 12247101), the Natural Science Foundation of Jiangsu Province (Grant No. BK20220876), the Fundamental Research Funds for the Central Universities (Grant No. lzujbky-2024-jdzx06), and the Natural Science Foundation of Gansu Province (Grant No. 22JR5RA389).
The computations are partially supported by High Performance Computing Platform of Nanjing University of Aeronautics and Astronautics.
\end{acknowledgments}

%

\end{document}